# Spatially resolved photoluminescence analysis of Se passivation and defect formation in CdSe$_x$Te$_{1-x}$ thin films

A. R. Bowman[1,2,3*], J. F. Leaver[4*], K. Frohna[2,3], S. D. Stranks[2,3], G. Tagliabue[1+] and J. D. Major[4+]

1. Laboratory of Nanoscience for Energy Technologies (LNET), STI, École Polytechnique Fédérale de Lausanne (EPFL), Lausanne 1015, Switzerland
2. Cavendish Laboratory, Department of Physics, University of Cambridge, J.J. Thomson Avenue, Cambridge, CB3 0HE, UK
3. Department of Chemical Engineering & Biotechnology, University of Cambridge, Philippa Fawcett Drive, Cambridge, CB3 0AS, UK
4. Stephenson Institute for Renewable Energy, University of Liverpool, L69 7ZF, UK

*These authors contributed equally to this work
+corresponding authors: giulia.tagliabue@epfl.ch & jonmajor@liverpool.ac.uk

*Abstract*

CdTe is the most commercially successful thin-film photovoltaic technology to date. The recent development of Se-alloyed CdSe$_x$Te$_{1-x}$ layers in CdTe solar cells has led to higher device efficiencies, due to a lowered bandgap improving the photocurrent, improved voltage characteristics and longer carrier lifetimes. Evidence from cross-sectional electron microscopy is widely believed to indicate that Se passivates defects in CdSe$_x$Te$_{1-x}$ solar cells, and that this is the reason for better lifetimes and voltages in these devices. Here, we utilise spatially resolved photoluminescence measurements of CdSe$_x$Te$_{1-x}$ thin films on glass to study the effects of Se on carrier recombination in the material, isolated from the impact of conductive interfaces and without the need to prepare cross-sections through the samples. We find further evidence to support Se passivation of grain boundaries, but also identify an associated increase in below-bandgap photoluminescence that indicates the presence of Se-enhanced luminescent defects. Our results show that Se treatment, in tandem with Cl passivation, does increase radiative efficiencies. However, the simultaneous enhancement of defects within the grain interiors suggests that although it is overall beneficial, Se incorporation may still ultimately limit the maximum attainable efficiency of CdSe$_x$Te$_{1-x}$ solar cells.

*Introduction*

CdTe-based solar cells are currently the only commercial competitor to silicon cells, and offer the lowest levelized cost of electricity (LCOE) of any photovoltaic technology for utility-scale generation[Zidane2019]. However, the record power conversion efficiency for a small-area CdTe cell is 22.3%, markedly lower than the 26.8% for crystalline silicon[Green63] despite the higher theoretical maximum efficiency for CdTe[Ruhle2016]. As such, there is significant scope to improve CdTe device efficiencies and further drive down the LCOE of solar electricity generation.

A relatively recent innovation in CdTe-based solar cells has been the introduction of a Se-alloyed CdSe$_x$Te$_{1-x}$ region at the front (light-facing side) of the absorber, with a graded composition from higher

Se content at the front towards pure CdTe at the back of the device. The CdSe$_x$Te$_{1-x}$ region has a lower bandgap than CdTe for moderate Se content[Tit2009,Lingg2018], resulting in a bandgap grading in these devices (from lower at the front to higher at the back). The lower bandgap at the front of CdSe$_x$Te$_{1-x}$ device extends light absorption further into the infrared, increasing the photocurrent compared to pure CdTe devices[Paudel2014,Ablekim2019]. Surprisingly, CdSe$_x$Te$_{1-x}$ devices have been demonstrated with similar open circuit voltages ($V_{OC}$) to pure CdTe devices[Munshi2018,Swanson2017], indicating a reduction in $V_{OC}$ losses that compensate for the lower bandgap. Cathodoluminescence (CL) measurements have shown increased luminescence, especially at grain boundaries, associated with higher Se content[Fiducia2019,Fiducia2022], suggesting that Se passivates defects in CdSe$_x$Te$_{1-x}$ devices, leading to reduced $V_{OC}$ losses. However, cross-sectional CL measurements can be influenced by surface effects as a result of the need to create an exposed cross-section through the device via mechanical or ion beam polishing, as well as generating charges in significantly different intensities and regions of the device and location compared to optical illumination. Photoluminescence (PL) thus offers important complementary evidence of carrier recombination in intact material.

Bulk photoluminescence measurements were recently employed on CdSe$_x$Te$_{1-x}$ devices to explore the role of Se[Kuciauskas2023]. Specifically, temperature dependent PL suggested that, for some fabrication processes, samples containing Se had sub-bandgap defects that limited the maximum achievable $V_{OC}$. Similar sub-bandgap PL defects have also been observed by Hu et al.[Hu2023] in Se containing samples at low temperatures. In both cases the nature of this defect and its relationship to sample morphology were not discussed. Microscale PL is a technique that can begin to answer these questions. It has previously been used to assess charge transport and passivation in both CdTe and CdSe$_x$Te$_{1-x}$ samples separately[Abudulimu2022, Alberi2013], and to explore the relationship of defects and grain boundaries to sample morphology in pure CdTe films[Liu2014] and CdSe$_x$Te$_{1-x}$ in giant grain samples[Neupane2023]. Similarly, time-resolved and spatially-resolved PL measurements have revealed heterogeneities in charge lifetimes in CdSe based solar cells[Xue2023], and two photon PL microscopy has also been helpful in disentangling surface and bulk transport and recombination effects in CdTe samples[Kuciauskas2017, Barnard2013], including passivation effects of dopants[Kuciauskas2016, Johnston2015] and in CdSe$_x$Te$_{1-x}$ samples[Kuciauskas2021]. Notably, prior microscale PL studies have measured samples in full or partial device stacks. This presents a fundamental issue: it is impossible to disentangle spectroscopically whether changes in charge lifetime or luminescence strength originate from charge extraction or sample passivation processes. Furthermore, no study has investigated the role of Se in samples in a controlled manner with CdSe concentration being carefully varied between samples. Finally, to date nearly all studies have focused on relative values, rather than absolute quantum efficiencies and absorption – the key complementary measurement to PL – has not been recorded at the microscale.

Here we fabricated CdSe$_x$Te$_{1-x}$ thin films on glass by interdiffusion of separately deposited CdSe and CdTe layers, with varying thicknesses of CdSe from 0 nm to 200 nm to alter the Se content in the final film. We are able to assess the role of Se on luminescence and recombination by measuring the spatially resolved absorption and photoluminescence of the samples. Fabrication on glass also allows us to directly assess the impact on the absorber layer, separated from the remainder of the device structure. Our results show that Se is distributed throughout the films and clearly passivates grain boundaries, in line with previous cross-sectional electron microscopy studies[Fiducia2019,Guo2019]. While Se alloying is shown not to significantly impact on the disorder of the main luminescence peak (as found from Urbach tails), we find it does increase the strength of below-bandgap defect luminescence. Our results suggest that these defect states are more present in grain interiors, implying a distinction in Se behaviour between grain boundary and bulk. This work provides further support for grain boundary passivation by Se, but crucially also indicates that the effects of Se might not be wholly beneficial, introducing defect states which could provide an ultimate cap to attainable $V_{oc}$.

*Methods*

Sample fabrication: CdSe$_x$Te$_{1-x}$ films with varying Se content were prepared on glass, without any conductive layers, in order to eliminate the effects of carrier extraction on the luminescence results. Soda-lime glass microscope slides were cut to 2.5 cm by 2.5 cm squares and cleaned by: scrubbing with a 1% Hellmanex™ solution by volume, a 15 min ultrasonic bath in de-ionised H$_2$O, then sequential rinses in de-ionised H$_2$O, acetone and propan-2-ol. This was followed by 10 min in an Ossila UV Ozone Cleaner to remove remaining organics from the surface. Various thicknesses of CdSe were deposited on the substrates by radio-frequency sputtering with a power density of 1.32 W/cm$^2$, Ar pressure of 5 mTorr and a substrate temperature of 200°C. CdTe was deposited by close-space sublimation (CSS), with a deposition pressure of 5 Torr in an N$_2$ atmosphere and a source temperature of 600°C. A 20 min post-growth anneal was performed in the CSS chamber at 600°C and 400 Torr immediately after deposition. A Cl treatment was performed on some of the samples by placing them in a tube furnace adjacent to glass plates coated with 1 mol/dm$^3$ aqueous MgCl$_2$ solution, and annealing in air at 410°C for 20 min. Excess chloride was rinsed off with de-ionised H$_2$O.

The CdSe and CdTe layers, whilst deposited separately, are expected to interdiffuse during the post-growth CSS anneal and Cl treatment steps, resulting in a CdSe$_x$Te$_{1-x}$ film[Ablekim2020]. It was found that substrate adhesion was particularly poor for the films deposited on glass. As such, a standard Cl treatment, involving drop-coating the MgCl$_2$ solution onto the samples before annealing, resulted in unacceptable levels of sample delamination from the substrate. The Cl treatment approach used in this work reduced delamination issues whilst still providing some Cl treatment.

Absorption: measurements (referred to as the 'second measurement system' in the text) were recorded on an NT&C system (a Nikon Eclipse Ti2 coupled to a Princeton Instruments Spectra Pro HRS-500 spectrometer and a PIXIS 256 camera). The light source for transmission measuremnts was a Halogen bulb and for reflection an Energetica LDLS$^{TM}$ laser driven white light source output through a fibre with 100 μm core diameter. For transmission incident light was focused on the sample through a condenser lens with variable a-stop (almost fully closed), while for reflection incident light was focused in the centre of the objective back focal plane. In both cases this allowed light to be incident perpendicular to the sample. All recorded signals were passed through a 450 nm long pass filter (Thorlabs FELH0450) to prevent second order effects on the spectrometer. For transmission a bare quartz substrate was used as a reference, while for reflection a mirror of known spectral response (Thorlabs PF10-03-P10-10) was employed as the reference (the Thorlabs reported spectral response was used).

Photoluminescence and photoluminescence quantum yield (PLQY): was carried out on a Renishaw inVia Raman Microscope RE04 using a Coherent sapphire 488 SF NX excitation, incident on the air-$CdSe_xTe_{1-x}$ interface as the thickness of the glass substrate prevented good focus on the glass-$CdSe_xTe_{1-x}$ interface. All maps were recorded via snake scanning of the diffraction limited spot across the sample. We note the sample stage moved rather than the spot in all cases. PLQY was estimated using a similar approach to that outlined by[Frohna2022]. Specifically, a calibrated light source (Ocean Optics HL-3P-INT-CAL) was coupled to an integrating sphere with known spectral response (Thorlabs 2P3/M). The output port of this sphere was aligned with the objective lens (Lecia N PLAN EPI, 100x objective, NA=0.85) of the Raman microscope and the spectrum recorded (allowing for relative radiometric calibration of PL data). The notch filter that normally removes the laser signal from the recording path was then removed, and the signal again recorded. The ratio of these signals gives the spectral response with and without the notch filter present. Secondly, the sphere was removed and the objective lens focused on a mirror (Thorlabs PF10-03-P10-10) with known spectral response. The laser normally used to excite the sample was then shone on the mirror (at low power) and the signal recorded on the spectrometer. Finally, the mirror is removed and the incident power at the same position recorded with a Thorlabs S170C or S130C power meter. By accounting for the reflection strength of the mirror, the ratio of number of photons incident on the mirror and the number recorded by the spectrometer can be calculated i.e. an absolute calibration at one wavelength. Using the recorded lamp spectrum, it is then possible to work out an absolute calibration at any wavelength both with the notch filter present and removed. This calibration allows for absolute luminescence yield measurements. We note that the relative radiometric response of the sphere was measured by sending a collimated white light beam into a second integrating sphere and recording the signal at the output port on an Ocean Insight Spectrometer FLAME-S-XR1. The sphere of interest was then coupled to the second integrating sphere and the collimated beam sent into the sphere of interest. The ratio of these two signals gives the spectral response of the integrating sphere used in measurements. An incident laser intensity of $5.9 \times 10^6$ mWcm$^{-2}$ was used in all measurements unless otherwise stated (noting this was a beam focused in a small area that we assume diffraction limited). Spot size here is defined as the boundary through which the intensity falls to $\frac{1}{e^2}$ of the maximum intensity. Intensity of the laser was selected to allow for rapid sample measurement while preventing sample degradation (as observed by a gradual drop in the sample's photoluminescence with time, which was seen at higher laser intensities). When calculating PLQY we assume Lambertian emission into a full $\pi$ hemisphere forwards and backwards, that is we multiply the number of recorded photons by $\frac{2}{NA^2}$, where $NA$ is the numerical aperture of the objective, equal to 0.85 in these measurements. Furthermore, we were unable to measure absorption from the same region as PL, so when calculating PLQY we used the spatially averaged absorption value of 0.9 (see Figure S3). We also note that we initially attempted PL measurements using wide-field illumination and hyperspectral mapping, but due to CdTe's weak PL signal usable data was not obtained.

Atomic force microscopy (AFM): was used to measure sample thicknesses. A Bruker FastScan AFM was employed for all measurements in ScanAsyst mode. ScanAsyst-Fluid+ tips were used in all measurements. 1st order plane fits were applied to all results. We note in two AFM scans (out of more than 16 scans in total) we observed a large number of very small (<1 μm size) grains nestled between larger grains. We removed these scans prior to the roughness analysis presented in the Supporting Information.

*Main*

CdTe thin films were deposited either directly onto glass substrates or with varying thicknesses of CdSe underlayer present (intermixing of a CdSe/CdTe bilayer to form $CdSe_xTe_{1-x}$ is conventional in solar cell fabrication), both with and without post deposition Cl passivation. Cl passivation was included as an additional variable due to its known influence on grain boundary passivation[Major2016,Moseley2014] and its ability to modify interdiffusion[Potter2000]. Simple glass substrates, as opposed to full device stacks, allowed us to directly examine the passivation properties of Se in $CdSe_xTe_{1-x}$ thin films in isolation. Cell structure samples (i.e. fabricated on typical electron/hole transport layers) exhibiting strong luminescence can be indicative of either better passivation or weaker charge extraction. Thus, producing samples on glass allows a cleaner analysis of passivation. This sample structure is not without challenge due to the lack of surface roughness typically provided by underlying layers. Following chloride treatment, samples were found to have a degree of non-uniformity on the microscale. Specifically, the chloride treatment resulted in delamination of some sample regions from the substrate while others were left intact. This is due to the low roughness of the glass substrate, compared to typical underlying layers, limiting the adhesion of the CdTe film during recrystallisation which occurs during Cl treatment. As an example of delamination, Figure 1a shows a white light reflection image of a Cl-treated pure CdTe sample, i.e. with a 0 nm CdSe underlayer. Due to the spatially resolved nature of the measurements used, analysis was able to focus solely on areas where grains remain intact. We additionally note that, in none delaminated regions, similar grain morphologies and surface roughnesses were observed for all chlorine-treated samples, as is shown in atomic force microscopy maps in Supporting Information Note 1. In all measurements presented, multiple positions were measured for all samples with representative results being presented in the main text and supporting information.

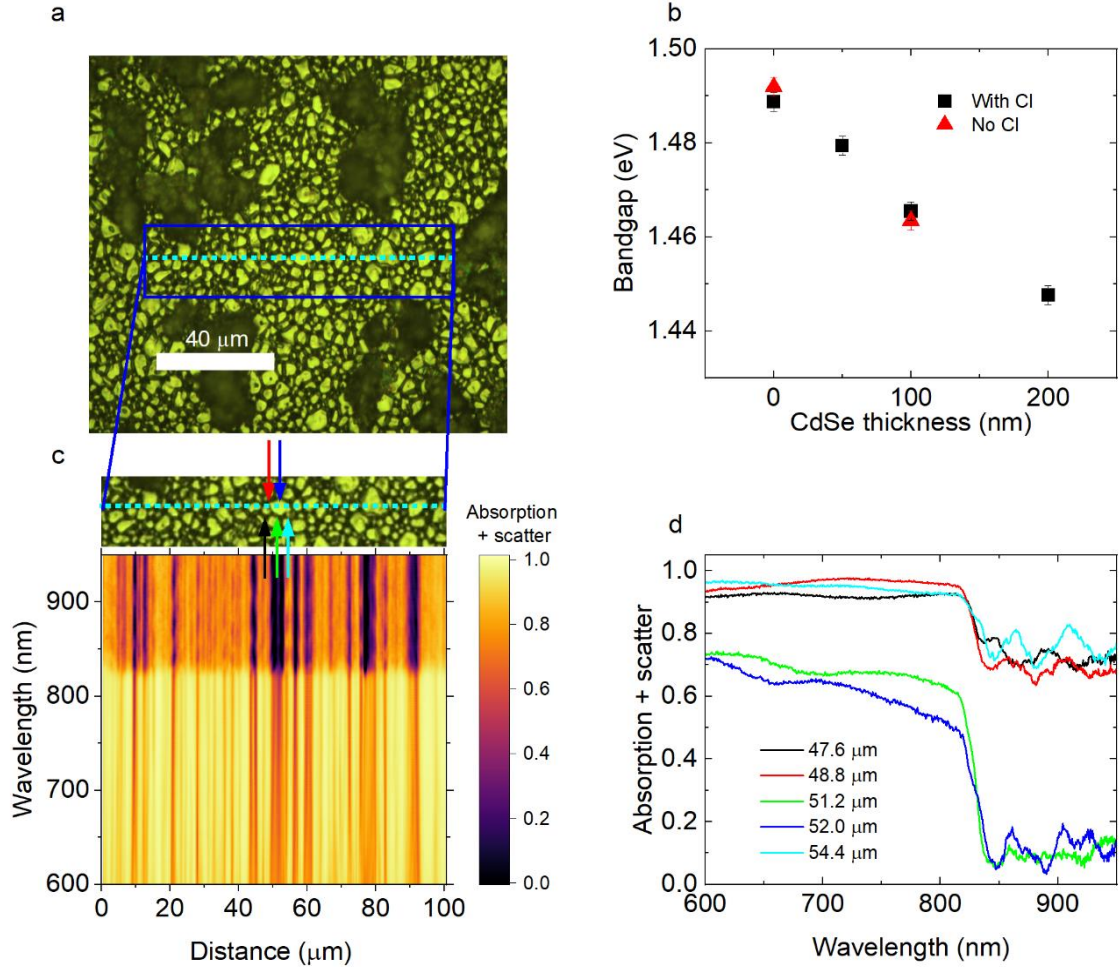

*Figure 1. a) white light reflection image of a Cl-treated 0 nm CdSe sample. Extended dark areas correspond to region of delamination. Overlaid on this plot is a lineslice along which we measured sample absorption. b) change in bandgap with CdSe thickness, via Tauc fitting of average absorption spectra from microscopic measurements, for samples both with and without a prior Cl-treatment. c) absorption + scatter results obtained for a 0 nm CdSe, Cl-treated sample. Above the main plot is the white light image of the sample with a blue dashed line highlighting the region measured. d) absorption + scatter for specific distances, as marked with colour arrows for specific positions in c), as the measurement moves across a grain. Weaker below-bandgap scattering is seen when measuring on a grain (corresponding to 51.2 μm and 52.0 μm).*

Microscale transmission and reflection measurements were carried out on all samples, which allowed us to calculate the fraction of light absorbed or scattered to high angles by the sample (equal to $1 - transmission - reflection$). Initially we used a hyperspectral imaging system to measure across a wide sample area (see Supporting Information Note 2). However, unexpectedly strong sub-bandgap inter-grain scattering was observed in all samples. To confirm whether these features were correct, we recorded the same signals along a lineslice (dashed line on Figure 1a) in a second measurement system (see Methods). While this second approach limits the number of points measured, it has previously reproduced absorption coefficients from single crystals to an extremely high level of accuracy[Bowman2023]. We found extremely similar results in both cases, confirming strong sub-bandgap scattering (discussed further below). We present further results from the second measurement approach in the main text.

The deposited CdSe layers were expected to completely interdiffuse with CdTe, resulting in $CdSe_xTe_{1-x}$ films, with higher Se content for thicker CdSe layers. We carried out Tauc plot fitting (with sub-bandgap scattering subtracted) of the spatially averaged optical absorption data (see Supporting Information Note 3) to determine the average bandgap of each sample[Tauc1968]. The bandgap reduced with increased CdSe thickness, both with and without prior Cl treatment, as is presented in Figure 1b. This demonstrates that samples behave as expected and we reproduce previous literature results well[Lingg2018].

A spatially and spectrally resolved absorption map for a Cl-treated 0 nm CdSe sample is presented in Figure 1c, alongside an image marking the sample region measured (for other samples see Supporting Information Note 3). Above the bandgap, strong absorption is observed at all wavelengths and positions, as expected for any good solar cell material. For grain centres absorption is never stronger than 0.75, which is reasonable noting the air-CdTe Fresnel power reflection coefficient is approximately 0.25. Below-bandgap (> 820 nm) scattering reaches values greater than 0.5 between some grains but is close to 0 at grain centres. This is further demonstrated in Figure 1d, where *absorption + scatter* is plotted for specific positions across a grain – high below-bandgap values are seen only between grains. Therefore, inter-grain regions scatter light into angles higher than 44° (based on the microscope objective NA of 0.7). We found that the extent of this inter-grain scattering was similar in all samples (Supporting Information Note 3), implying scattering is not the main cause of changes to luminescence observed in subsequent parts of this work.

We attempted to use our absorption measurements to probe for bandgap changes across the sample – specifically whether we could correlate higher quantities of Se with lower bandgaps at grain boundaries. This was done by subtracting below-bandgap scattering from all samples and performing a Tauc fit at each position individually (see Supporting Information Note 4). Larger bandgap variation was observed for samples containing more Se, and the data is suggestive of lower bandgaps near gran boundaries. However, due to spatial and noise resolution limits in our measurements, we were unable to draw definitive conclusions from absorption measurements alone.

To further explore the role of Se in our samples and explore passivation effects we employed confocal mapping to record PL across the sample surface. We present average PL spectra for all Cl-treated samples of various Se content samples, as well as for 0 nm and 100 nm CdSe untreated samples, in Figure 2a. We selected the incident laser power ($5.9 \times 10^6$ mWcm$^{-2}$) to be high enough to allow for efficient mapping, but still within a trap-dominated regime (see Supporting Information Note 5 for further discussion). For all samples a single main luminescence peak is seen. Interestingly, the PL peak wavelength of non-Cl treated samples shows no correlation with the quantity of CdSe present, despite Tauc fitting of the absorption measurements in Figure 1b showing a reduced bandgap for the non-Cl treated 100 nm CdSe sample. This leads us to suggest that, without Cl-treatment, either the CdSe$_x$Te$_{1-x}$ phase absorbs light but is non-emissive, or charge diffusion between regions richer in CdTe (that absorb the majority of the light incident on the sample, as this side is excited) and phases with higher Se content is minimal. In either case, only PL from the pure CdTe grains is observed. For Cl-treated samples the PL peak does shift to lower energies with increasing Se content, and the peak corresponds well with bandgaps estimated from Tauc fitting (Figure 2a, inset). In fact, for Cl-treated samples we were able to predict the sample luminescence well based on our measurements of sample absorption via the van Roosbroeck-Shockley relation[Roosbroeck1954], implying minimal Stokes shift prior to luminescence, as we discuss further in Supporting Information Note 6. For higher quantities of Se-incorporation in Cl-treated samples, the PL not only red-shifts, but the width of the peak also increases (full-width-half-maximum of 41 nm for 200 nm CdSe thickness, compared to 34 nm for 0 nm CdSe thickness). This again suggests that there is greater bandgap variation in samples with higher Se content. Disorder within the main luminescence peak, as measured by fitting the PL with an Urbach tail (see Urbach fits in Supporting Information Note 7)[Ledinsky2019], is relatively constant for all samples, varying between 10 meV and 13 meV at most, as is presented in Figure 2b. More surprisingly, as the quantity of Se is increased in Cl-treated samples, a longer, sub-bandgap wavelength (>900 nm) signal increases in relative intensity (see average photoluminescence on a logarithmic scale in Supporting Information Note 7). This indicates that as Se content is increased, a defect state with strong luminescence increases in strength. As already noted, similar effects have been observed in bulk measurements by Kuciauskas et al. and Hu et al.[Kuciauskas2023, Hu2023], but this is the first observation of a clear correlation between a controlled Se content and the strength of the defect luminescence.

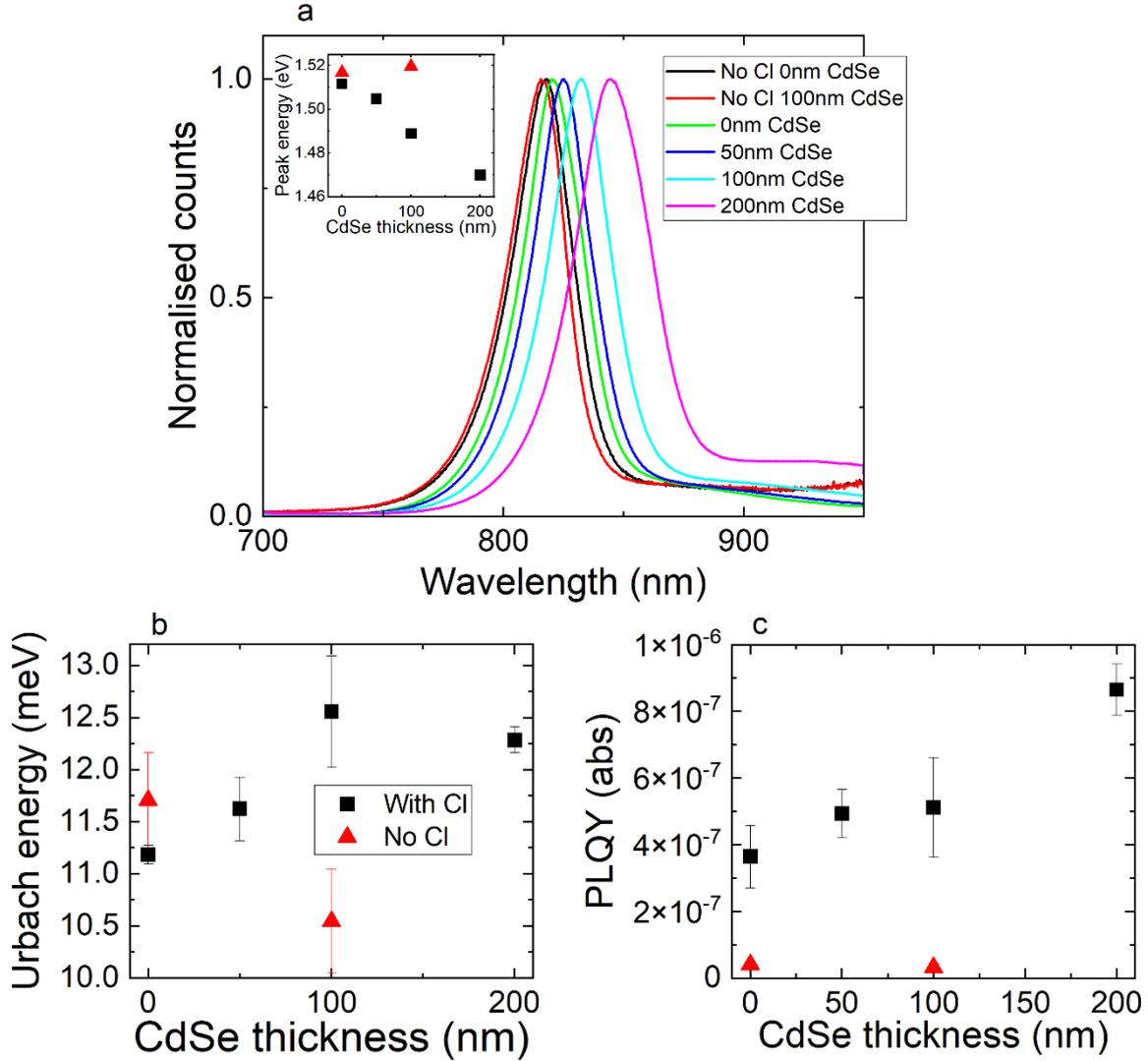

*Figure 2. a) spatially averaged photoluminescence, PL, observed from each map. The inset presents the peak PL position for each sample. b) spatially averaged Urbach energy for each sample, as a function of CdSe thickness. c) spatially averaged photoluminescence quantum yield (PLQY) as a function of sample thickness. Legend in b) applies to inset of a), b) and c).*

Figure 2c presents the spatially averaged photoluminescence quantum yield (PLQY) as a function of CdSe thickness. The PLQY measures of the ratio of radiative to total recombination rates – a perfect solar cell can only be achieved with a PLQY of unity. Here we observe PLQYs on the order of $10^{-7}$ - $10^{-6}$. Previous studies have reported PLQY values as high as $10^{-2}$ for surface-passivated $CdSe_xTe_{1-x}$ devices[Onno2022,Kuciauskas2020], although this drops to the order of $10^{-5}$ without surface passivation[Onno2022]. Similarly, the best open circuit voltges observed in full devices is ~ 0.9 V[Scarpulla2023,Mallick2023] while the detailed balance limit for CdTe predicts $V_{OC,ideal}$ =1.16 V[Geisthardt2015,Ruhle2016]. Noting $V_{OC} = V_{OC,ideal} + \frac{k_BT}{q}\ln(PLQY)$[Ross1967,Miller2012], this implies the best CdTe devices have PLQY ~ $4 \times 10^{-5}$. Thus the PLQY we observe is reasonable for samples without surface passivation that have been fabricated on glass, where surface recombination velocity would be expected to be higher than for an interface with optimised carrier extraction layers. Figure 2c confirms two additional points. Firstly, without Cl treatment the PLQY is approximately an order of magnitude lower, demonstrating this treatment is necessary for higher quality CdTe and that Se alone is not sufficient for passivation. Secondly the PLQY also increases with increased Se content (we observe an approximate doubling of PLQY by 200 nm of CdSe), demonstrating that, while it does not preclude the need for a Cl treatment, CdSe plays a key role in film passivation.

Having established the bulk $CdSe_xTe_{1-x}$ PL properties, we now consider microscale maps. Figure 3a-c/d-f shows a white light reflection image, peak wavelength map and local PLQY map for Cl-treated 0 nm/200 nm CdSe films. The peak wavelength is obtained by fitting the PL data from each pixel with a Gaussian, and details of PLQY measurement approach is provided in the methods section. Equivalent results for 50 nm and 100 nm CdSe films are provided in Supporting Information Note 8 along with maps for non-Cl treated films. We note previous microscale studies of CdTe based samples have not explored peak wavelength with position or directly quantified microscale PLQY [Neupane2023, Abudulimu2022, Liu2014, Alberi2013]. Our spatially resolved analysis allows direct comparison of emission properties across the grain structure thereby distinguishing between grain boundaries and grain interiors. For Cl-treated 0 nm CdSe, longer wavelength PL emission peaks (> 822 nm) are observed predominantly from grain centres (Figures 3a and b). As absorption measurements showed no significant bandgap variation across the sample (Supporting Information note 4), we interpret this to be a consequence of photon re-absorption i.e. shorter wavelength PL is re-absorbed before light can escape the grain, shifting the PL peak to longer wavelength from grain centres (as the PL must travel through more material before reaching the surrounding[Pazos-Outón2018]). A direct consequence would be a higher PLQY from grain boundaries, which is observed in Figure 3c, especially surrounding larger grains (for plots of overlaid grains and PLQY, see Supporting Information Note 9). Notably, we also see some grain-to-grain variation in the PLQY – some grains (especially in the bottom left of Figure 3c) are seen to have higher PLQYs. Grain-to-grain variation in PLQY is typical in many other semiconductor systems and can be related to numerous factors including strain and stoichiometry[Frohna2022].

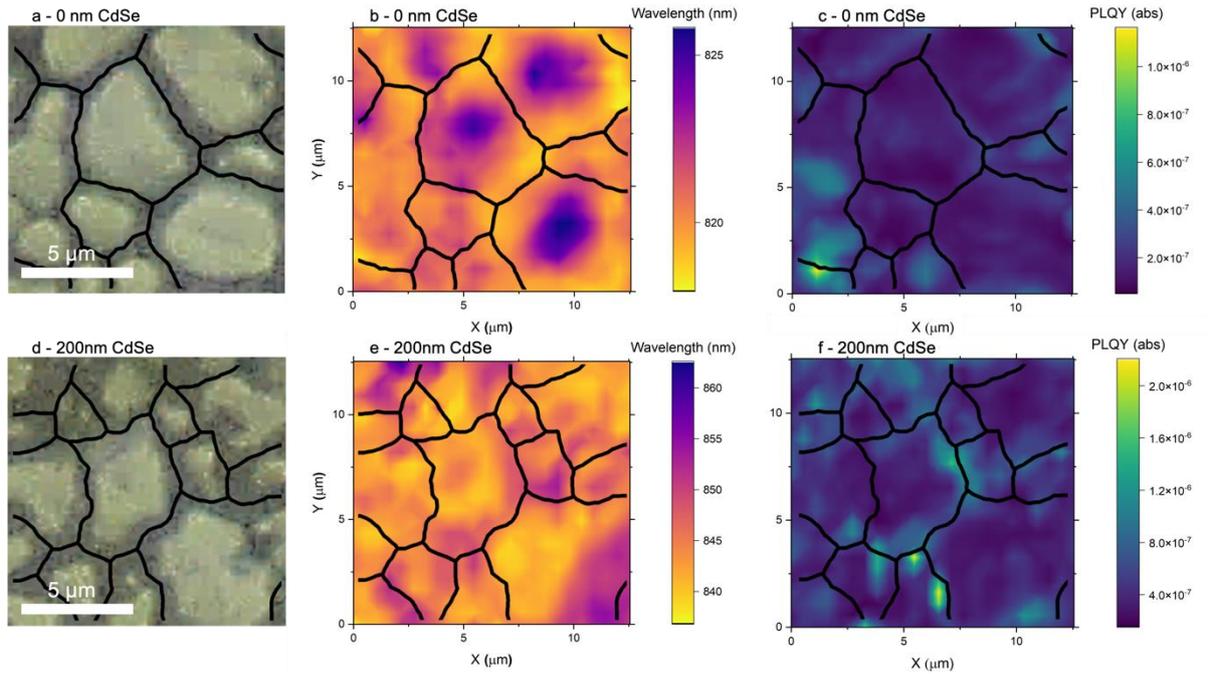

*Figure 3. a), b) and c)/d), e) and f) white light reflection, peak wavelength and peak PLQY plots (± 20 nm about the central peak) for 0 nm CdSe/200 nm CdSe samples, with both samples Cl treated. We note that the same colour scales are used for different ranges in b)/e) and c)/f) to allow for better visualisation of results. As a guide to the eye, black lines present approximate grain boundaries.*

The situation is completely reversed for Cl-treated 200 nm CdSe films. Here we observe longer peak wavelengths predominantly at grain boundaries and shorter peak wavelengths in grain interiors (Figure 3d/e, see Supporting Information Note 9 for overlays). This is the opposite of what we expect from photon re-absorption, so leads us to the conclusion (consistent with the greater variation of bandgap observed in absorption measurements) this must be a result of bandgap variation across the sample: grain edges must have lower energy bandgaps than grain centres, as has previously been observed in cross-sectional CL of $CdSe_xTe_{1-x}$ devices[Fiducia2019,Zheng2019]. We again note that surface roughness doesn't vary significantly between samples (Supporting Information Note 1), so this is not due to out-coupling alone. Additionally, we see significantly higher PLQYs from the grain boundaries compared to the grain centres, as is shown in Figure 3f. The variation in PLQY between grain boundary and grain centre is approximately a factor of 2 larger than for 0 nm CdSe, and as we found the scattering at grain boundaries to be comparable for both samples (Supporting Information Note 3), this reveals that the PLQY is stronger at grain boundaries due to better passivation. These results provide direct evidence that Se is incorporated at the grain boundaries and that it passivates these regions.

Cl-treated 50 nm CdSe and 100 nm CdSe plots show similar relationships between peak wavelength and PLQY to those presented in Figure 3, as shown in Supporting Information Note 8, with 50 nm CdSe behaving similarly to 0 nm CdSe and 100 nm CdSe more similarly to 200 nm CdSe. To further elucidate on this point (Supporting Information Note 10), a negative correlation (Pearson correlation coefficient, $r = -0.13$) is observed between peak wavelength and PLQY for the 0nm CdSe sample i.e. for longer wavelength regions (grain centres), the PLQY is lower (due to more photon re-absorption). However, as the quantity of CdSe is increased we observe a change to positive correlations ($r = 0.04, 0.18$ and $0.31$ for 50 nm, 100 nm and 200 nm CdSe respectively): longer wavelength regions have higher PLQYs. This again demonstrates that Se-rich regions (which have lower energy bandgaps) have better passivation. Due to much smaller grain sizes in non-Cl treated samples it was not possible to draw conclusions on the effect of grain structure on the PL, but we note we also see significant peak wavelength (± 12 nm) and PLQY variation (1 order of magnitude) across these samples

Finally, we return to the below bandgap PL that increases with thicker CdSe (c.f. Figure 2a) to understand its origin at the microscale. For this defect to be appropriately passivated, its location within the sample must first be understood. Some below bandgap PL was observed across nearly all sample regions (as shown in Supporting Information Note 11). To find where this emission was most pronounced relative to the main PL peak, we calculated PLQY maps of the long wavelength, below bandgap PL region (approximately 900 nm to 950 nm) and the tail of the main PL peak (specifically, the range 10-20 nm beyond the main PL peak, focusing on these wavelengths to remove photon re-absorption effects that would complicate analysis). Normalising these maps by their average value and dividing the resulting long-wavelength data by the main PL peak tail data gives us a ratio that, if larger than one, indicates a region with particularly strong below-bandgap PL relative to its emission at the main PL peak wavelength. This ratio and associated white light reflection images for Cl-treated 0 nm CdSe/200 nm CdSe are shown in Figure 4a-b/c-d. For 0 nm CdSe we observe a weak correlation between increased below-bandgap PL and grain boundaries (Figure 4b), and the ratio stays between 0.7 and 1.3 for all positions, indicating only a modest difference between grain interiors and grain boundaries. In contrast, for 200 nm CdSe we see a strong increase in long wavelength PL (with the ratio reaching values larger than 2), mainly focused at grain centres. This leads us to the conclusion that while Se alloying passivates grain boundaries, its presence also correlates with an increase in an emissive defect state more concentrated within the grain centres, which would ultimately be detrimental to solar cell performance. Hence these results imply that while Se incorporation helps to compensate for the deleterious impact of grain boundaries within the material, it may hinder further improvements by enhancing the formation of a luminescent trap state within the grains.

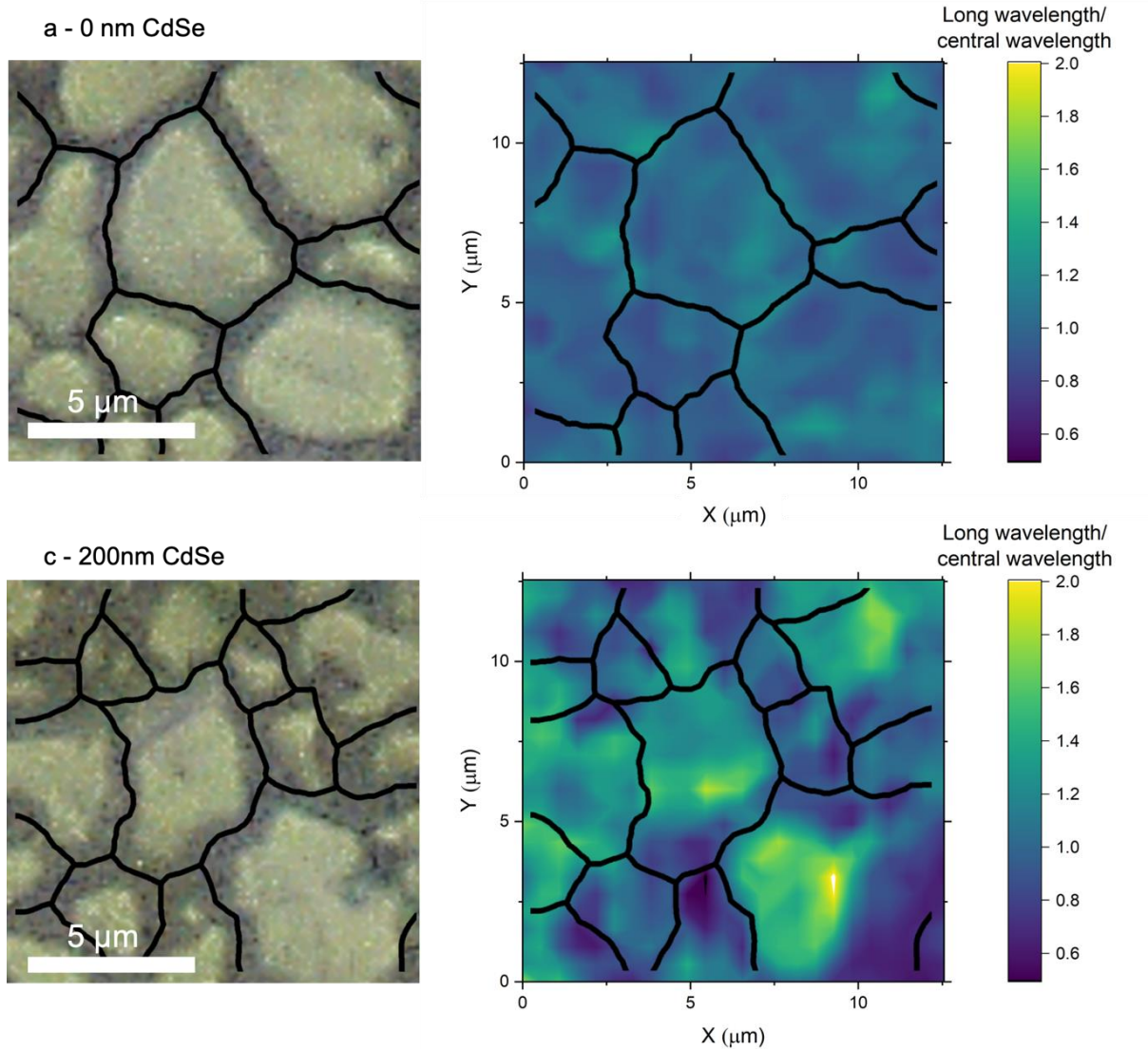

*Figure 4. a)/c) white light reflection and b)/d) long wavelength PL intensity normalised to average value divided by the main wavelength PL intensity normalised to average value (see main text), for Cl-treated 0 nm/200 nm CdSe. As a guide to the eye, black lines present approximate grain boundaries.*

*Conclusion*

This work showcases the possibilities of fully quantatative spatially resolved steady-state spectroscopy to understand CdTe and $CdSe_xTe_{1-x}$, providing new insight into the role of Se in incorporation in $CdSe_xTe_{1-x}$. A tailored series of $CdSe_xTe_{1-x}$ films on glass were studied, where the Se content was varied via controlled thickness of a CdSe underlayer. Our microscale absorption and photoluminescence measurements give direct evidence that higher Se content passivates grain boundaries in $CdSe_xTe_{1-x}$, corroborating prior results from cross-sectional CL studies. We further reveal that Se alloying leads to the increase of luminescent sub-bandgap states, which are more strongly concentrated in grain centres. Our results indicate that the defect chemistry associated with Se is more complex than has been previously appreciated and likely has both beneficial and detrimental effects on the absorber material. Further work is now required to understand the precise nature of this defect stateand whether it's formation can be avoided by modification of the fabrication process.

*Acknowledgements*

ARB acknowledges support of an EPSRC Impact Accelerator Grant, SNSF Eccellenza Grant PCEGP2-194181 and an SNSF Swiss Postdoctoral Fellowship TMPFP2_217040. ARB thanks Franky Esteban Bedoya Lora for use of an Ocean Optics spectrometer. JFL and JDM acknowledge EPSRC for funding via EP/N014057/1, EP/W03445X/1 and the EPSRC Centre for Doctoral Training in New and Sustainable Photovoltaics EP/L01551X/1. KF acknowledges the support of an Engineering and Physical Sciences Research Council (EPSRC) Doctoral Prize Postdoctoral Fellowship and a Winton Sustainability Fellowship. SDS acknowledges support from the Royal Society and Tata Group (UF150033). The authors acknowledge the EPSRC (EP/R023980/1) for funding.

# Supporting Information: Spatially resolved photoluminescence analysis of Se passivation and defect formation in CdSe$_x$Te$_{1-x}$ thin films

A. R. Bowman[1,2,3*], J. F. Leaver[4*], K. Frohna[2,3], S. D. Stranks[2,3], G. Tagliabue[1+] and J. D. Major[4+]

1. Laboratory of Nanoscience for Energy Technologies (LNET), STI, École Polytechnique Fédérale de Lausanne (EPFL), Lausanne 1015, Switzerland
2. Cavendish Laboratory, Department of Physics, University of Cambridge, J.J. Thomson Avenue, Cambridge, CB3 0HE, UK
3. Department of Chemical Engineering & Biotechnology, University of Cambridge, Philippa Fawcett Drive, Cambridge, CB3 0AS, UK
4. Stephenson Institute for Renewable Energy, University of Liverpool, L69 7ZF, UK

*These authors contributed equally to this work
+corresponding authors: giulia.tagliabue@epfl.ch & jonmajor@liverpool.ac.uk

*Supporting Information Note 1 – Atomic Force Microscopy maps of samples*

We recorded Atomic Force Microscopy (AFM) maps of several regions of each sample to better understand their morphology, with representative maps presented in Figure S1. We recorded the standard deviation in height for each AFM map, a measure of the surface roughness. We found that surface roughness was comparable in all samples, with results presented in Table S1.

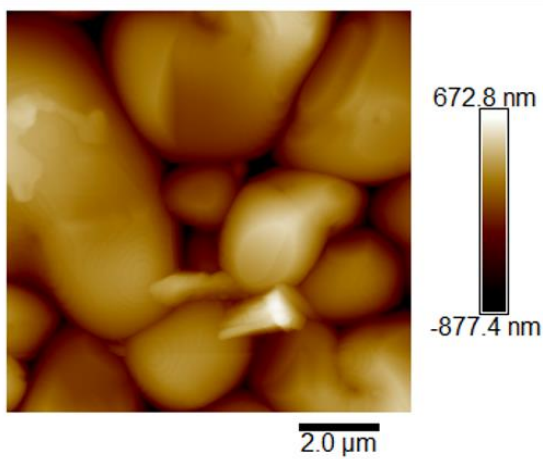
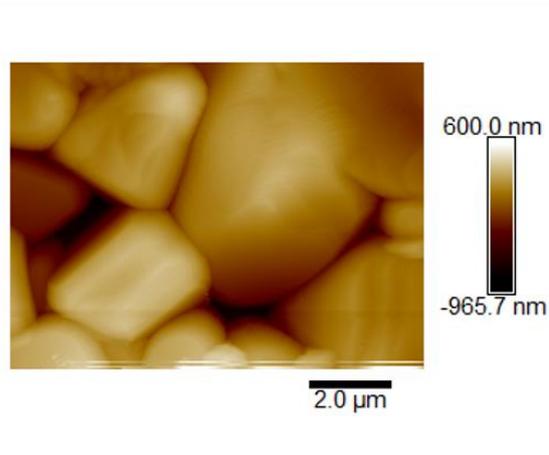
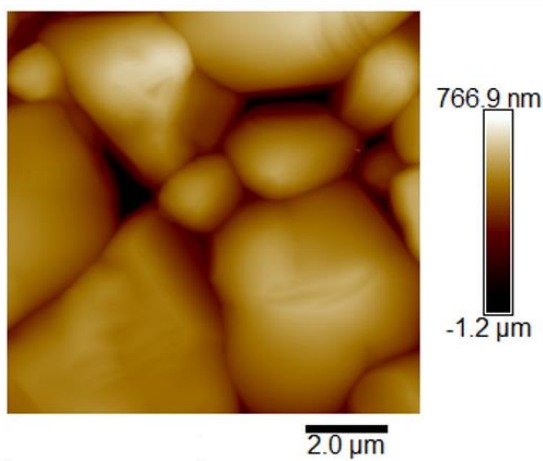
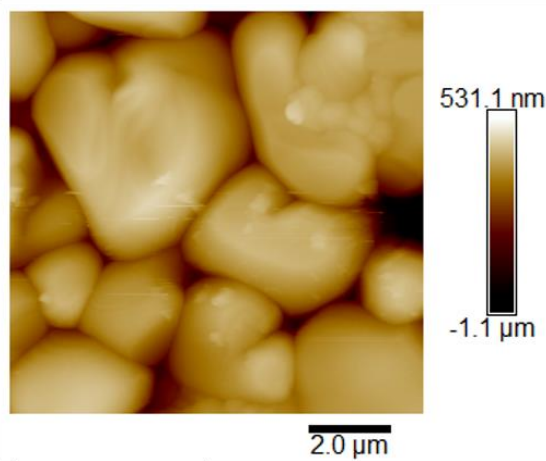

*Figure S1. Representative AFM maps of Cl-treated 0 nm/50 nm/100 nm/200 nm CdSe samples.*

| Sample (CdSe thickness, nm) | Average standard deviation in height, a measure of roughness, in AFM maps across samples measured (nm) | Standard deviation in roughness measurements (nm) |
|---|---|---|
| 0 | 186 | 17 |
| 50 | 176 | 35 |
| 100 | 178 | 50 |
| 200 | 214 | 7 |

Table S1. AFM results for Cl-treated $CdSe_xTe_{1-x}$ samples.

*Supporting Information Note 2 – hyperspectral absorption measurements*

We measured absorption across the sample also using a hyperspectral imaging system [1] and obtained similar results to those presented in the main text.

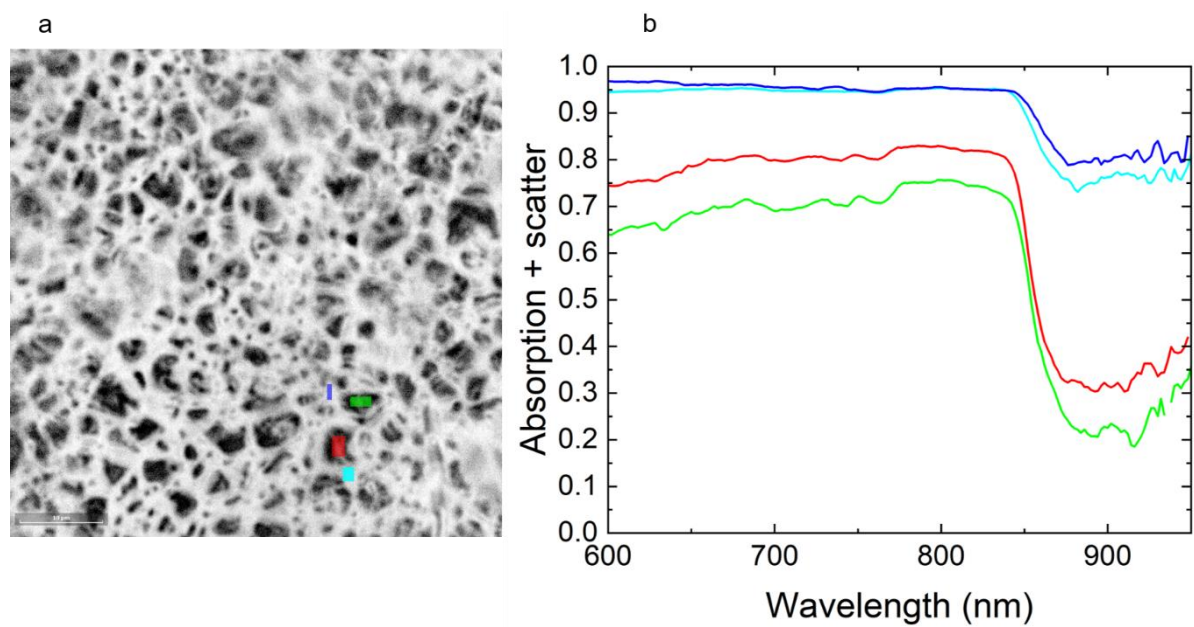

*Figure S2. a) presents the absorption + scattering at the band edge for a Cl-treated 200 nm CdSe film. b) presents the absorption + scattering for the specific coloured regions highlighted in a). It can be seen that very similar results are observed to those in Figure 1.*

*Supporting Information Note 3 – average absorption measurements*

Figure S3 presents the spatially averaged *absorption + scattering* for all samples. Figure S4 presents Tauc fitting of below-bandgap scattering subtracted data, while Figure S5 presents microscale *absorption + scattering* maps - the equivalent of Figure 1c - for other Cl-treated samples. It can be seen from these plots that, in the regions without grains present, the below-bandgap scattering is comparable for all samples.

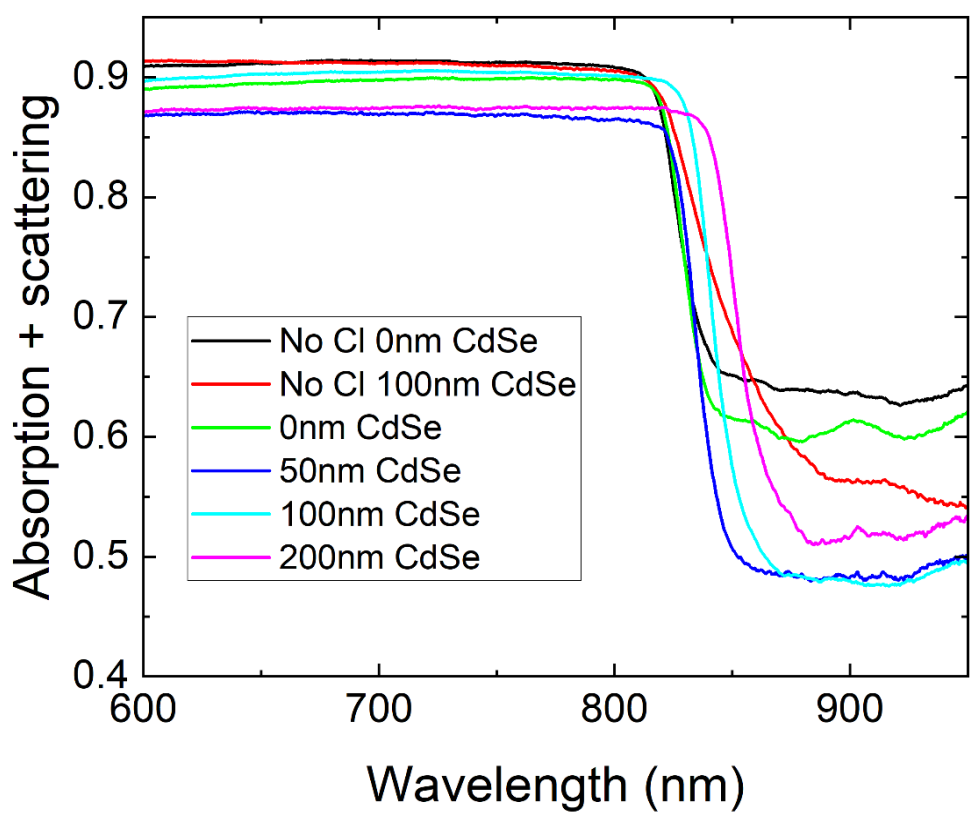

*Figure S3. Average absorption for all samples for measurement series presented in the main text.*

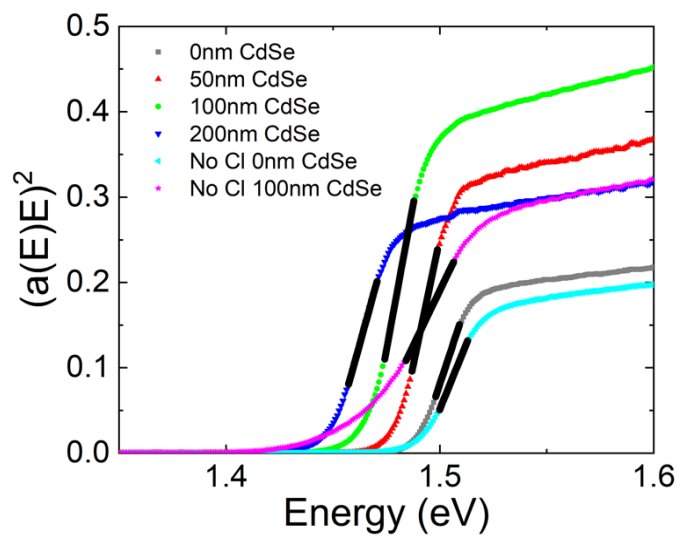

*Figure S4. Tauc fits (black lines) to scattering subtracted absorption data.*

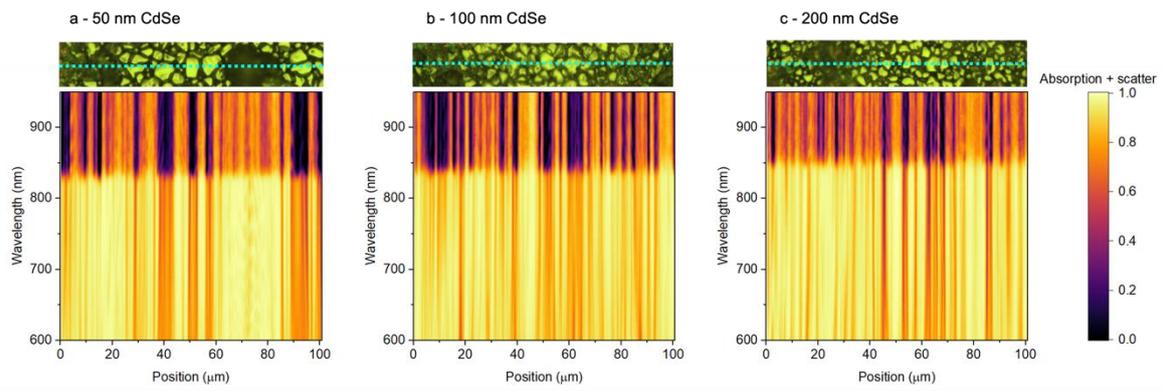

*Figure S5. Absorption + scattering maps for Cl-treated 50 nm, 100 nm and 200 nm CdSe are presented in a), b) and c) respectively.*

*Supporting Information Note 4 – Tauc fitting absorption measurements*

We carried out Tauc fitting of spatially resolved absorption measurements recorded for both lineslice measurements (Figure S6) and the hyperspectral imaging system (Figure S7). In the case of the lineslice, more variation in bandgap is observed for the (Cl-treated) 200 nm CdSe sample than the 0 nm CdSe sample, and decreases in bandgap roughly correlate with grain boundaries. However, spatial resolution in this system prevents us from drawing definitive conclusions. In the hyperspectral approach we again observe more bandgap variation for 200 nm CdSe (see Figure S7d compared to Figure S7c). However, here the below-bandgap data was extremely noisy, especially between grains. Thus we were not able to carry out Tauc fitting at all regions of the map, again preventing us from drawing definitive conclusions. In both cases the data is suggestive of lower bandgaps at grain boundaries.

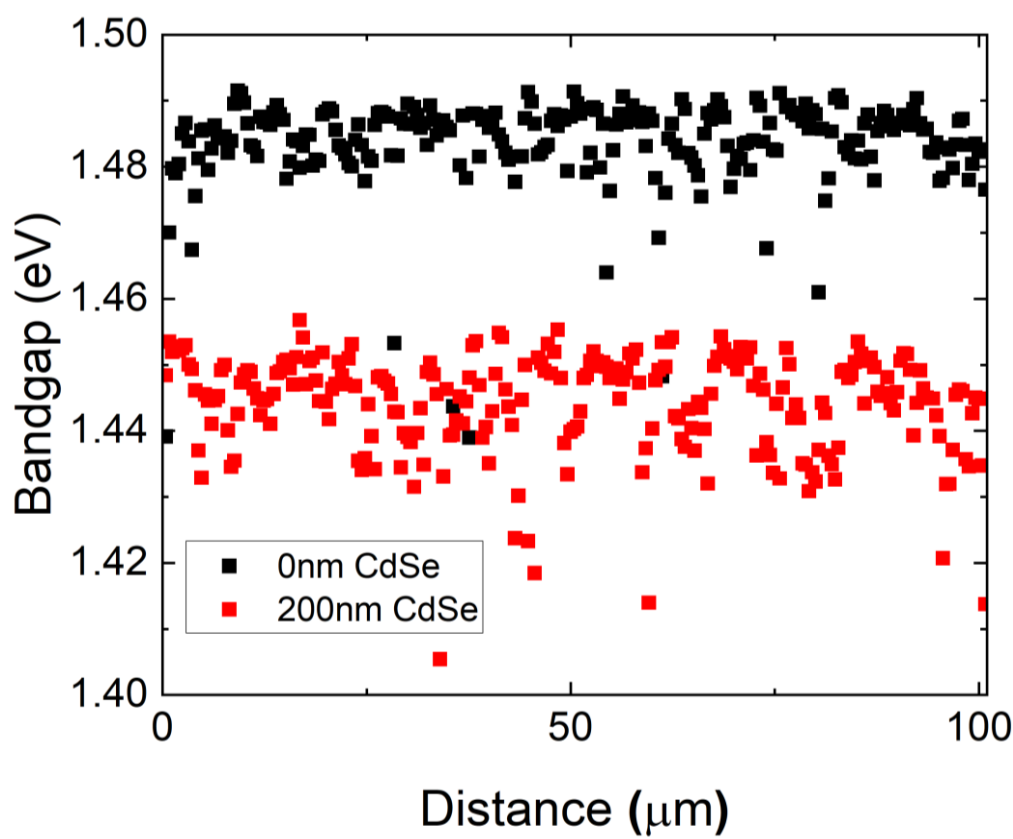

*Figure S6. The fitted Tauc bandgap at all positions across the lineslice measured for Cl-treated 0 nm CdSe and 200 nm CdSe. It can be seen there is more bandgap variation for the 200 nm CdSe.*

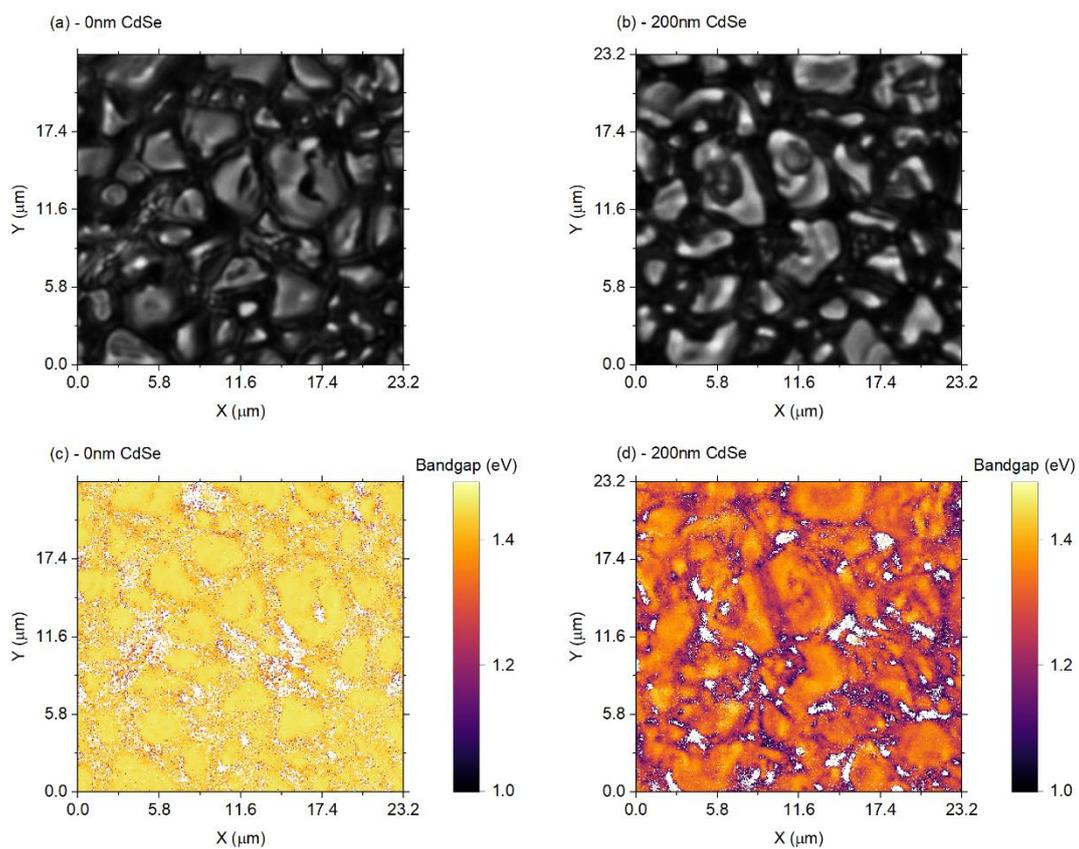

*Figure S7. a)/b) reflection images for 600 nm illumination and c)/d) bandgaps extracted from Tauc fits of absorption plots for Cl-treated 0 nm/200 nm CdSe. White regions of c) and d) are where data was too noisy to confidently extract a bandgap.*

*Supporting Information Note 5 – trap-dominated regime*

In steady state the local charge generation rate, $G$, is equal to the charge recombination rate $R$. Furthermore, for any position in the material $G \propto I$, the incident laser intensity, and $R \approx a(n - n_i) + b(n^2 - n_i^2) + c(n^3 - n_i^3)$, that is the recombination depends on first, second and third order recombination rates $a, b$ and $c$, where $n$ is the density of excited electrons and the subscript $i$ corresponds to intrinsic concentrations. We have assumed here that the density of excited electrons is approximately equal to the density of excited holes. Furthermore, we can state that (for an undoped system) the radiative recombination rate is $b_r(n^2 - n_i^2)$. Therefore, the recorded photoluminescence divided by the incident laser intensity is

$$\frac{PL_{recorded}}{I} \propto \frac{b_r(n^2 - n_i^2)}{a(n - n_i) + b(n^2 - n_i^2) + c(n^3 - n_i^3)} \approx \frac{b_r n}{a + bn + cn^2}$$

where for the last approximation we have assumed $n \gg n_i$, which is true for the excitation densities used in our experiments. Therefore, if we plot $\frac{PL_{recorded}}{I}$ versus $I$ and observe a positive gradient, this indicates we are operating in a trap-dominated regime where $R \approx an$ (as then $\frac{PL_{recorded}}{I} \propto I$). In Figure S8 we plot $\frac{PL_{recorded}}{I}$ versus $I$ for excitation densities of the same order of magnitude as those used in our mapping experiments and we observe an approximate straight line, confirming that we are within a trap-dominated regime.

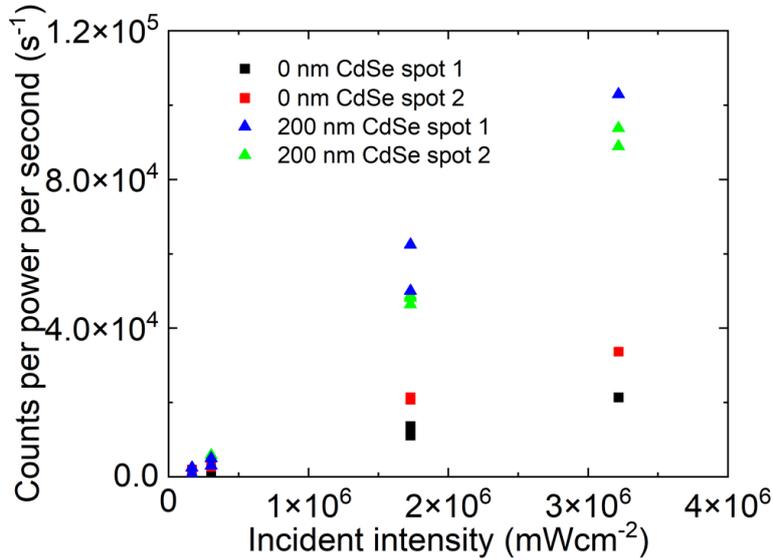

*Figure S8.* $\frac{PL_{recorded}}{I}$ *versus I for different points on 0 nm CdSe and 200 nm CdSe samples, with Cl treatment. Linear scaling is observed in all cases. Power was increased from minimum to maximum and then reduced back to minimum to search for hysteresis effects, hence multiple points at the same intensity for some samples.*

*Supporting Information Note 6 – prediction of photoluminescence from absorption measurements*

We applied the van Roosbroeck-Shockley relation [2] to predict the spectral shape of the photoluminescence from our absorption measurements. Specifically, this relation states that $PL(E) \propto a(E)E^2 e^{-\frac{E}{k_BT}}$, where $a(E)$ is the measured absorption as a function of energy $E$ and $k_BT$ the thermal energy. We used the scattering subtracted, spatially averaged $a(E)$ (i.e. that presented in Figure S4) and found we had to shift predicted values up by 8 meV (i.e. predicted PL peaks were at lower energies than measured PL peaks) to obtain good agreement with spatially averaged measured photoluminescence. We attribute this shift to a small calibration error between the two measurement systems used to record PL and absorption. Finally, we note that there is significant error in our below bandgap $a(E)$ measurements, so we only predict the energies close to the PL peak on the lower energy side. A comparison between experiment and theory is presented in Figure S9.

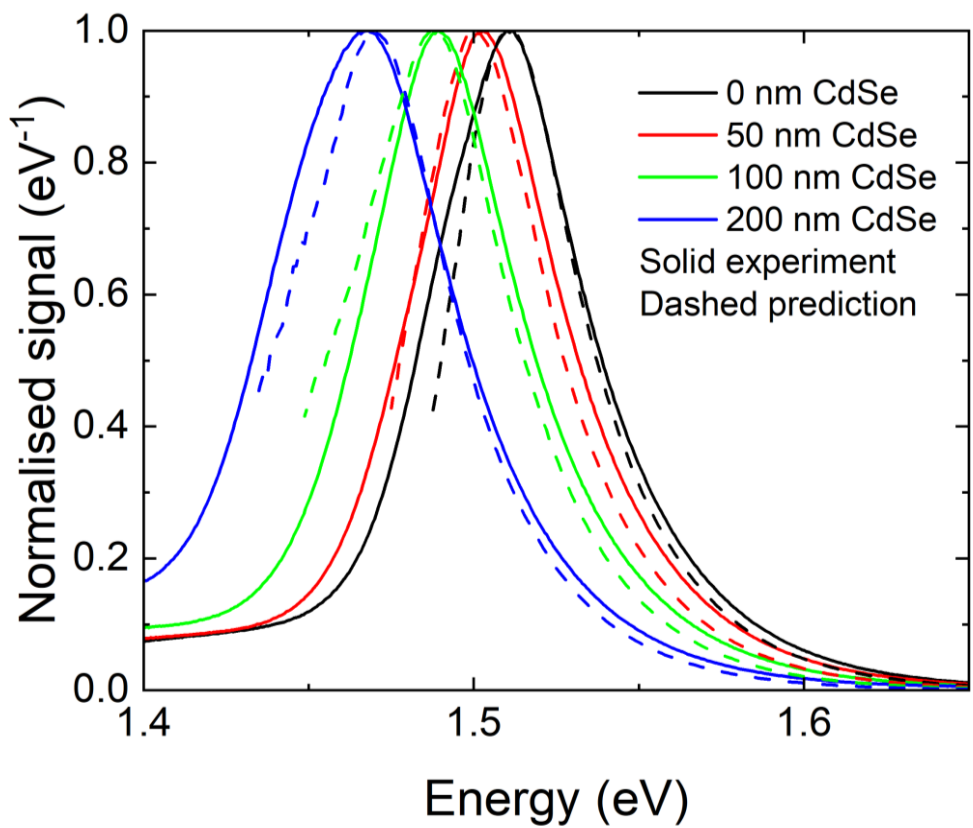

*Figure S9. Comparison between measured (solid lines) and absorption predicted (dashed lines) photoluminescence for Cl-treated samples. Predicted PL was shifted by 8 meV to have good agreement (see text above).*

*Supporting Information Note 7 – logarithmic scale spectra and Urbach fit*

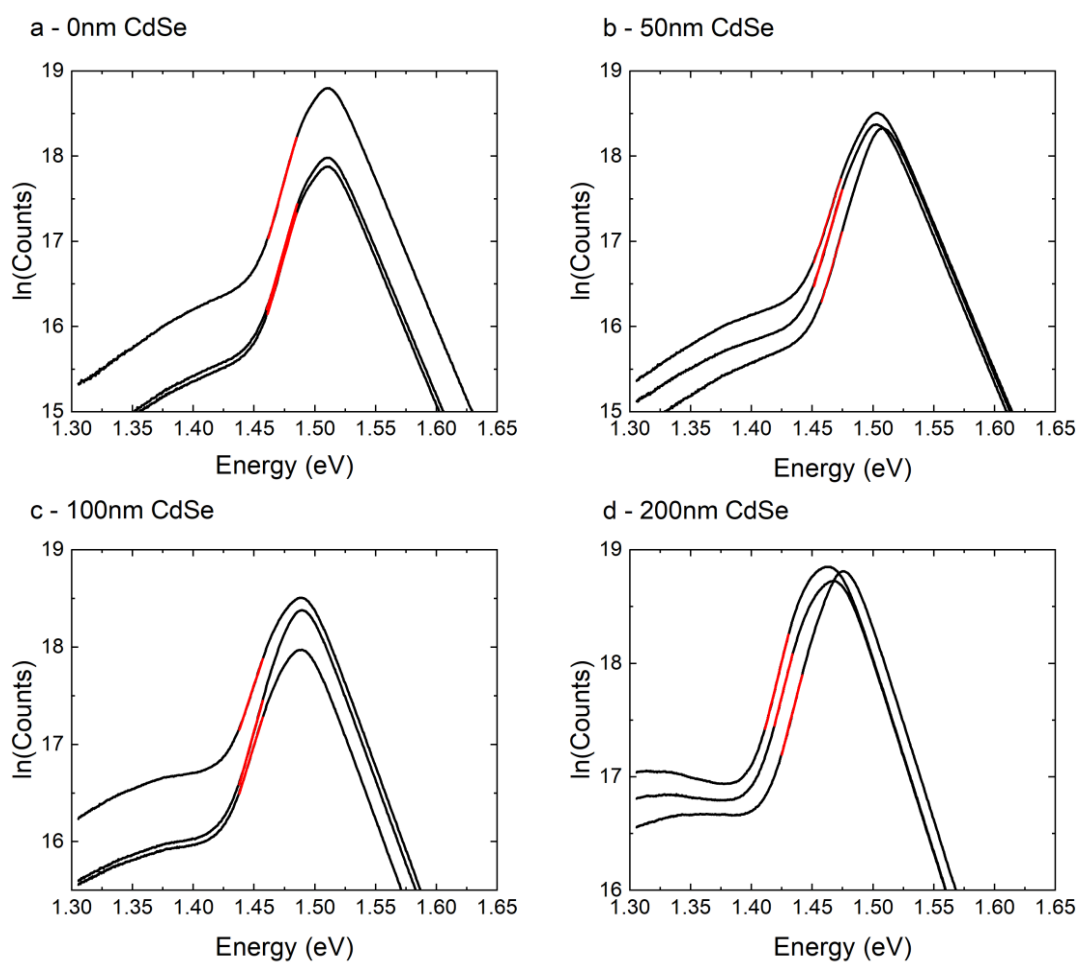

*Figure S10. a)/b)/c)/d) the natural logarithm of the spatially averaged photoluminescence spectrum from three regions mapped for 0 nm/50 nm/100 nm/200 nm CdSe as a function of energy (black). Overlaid on each plot is the Urbach fit to the below bandgap region (red).*

*Supporting Information Note 8 – other PL maps*

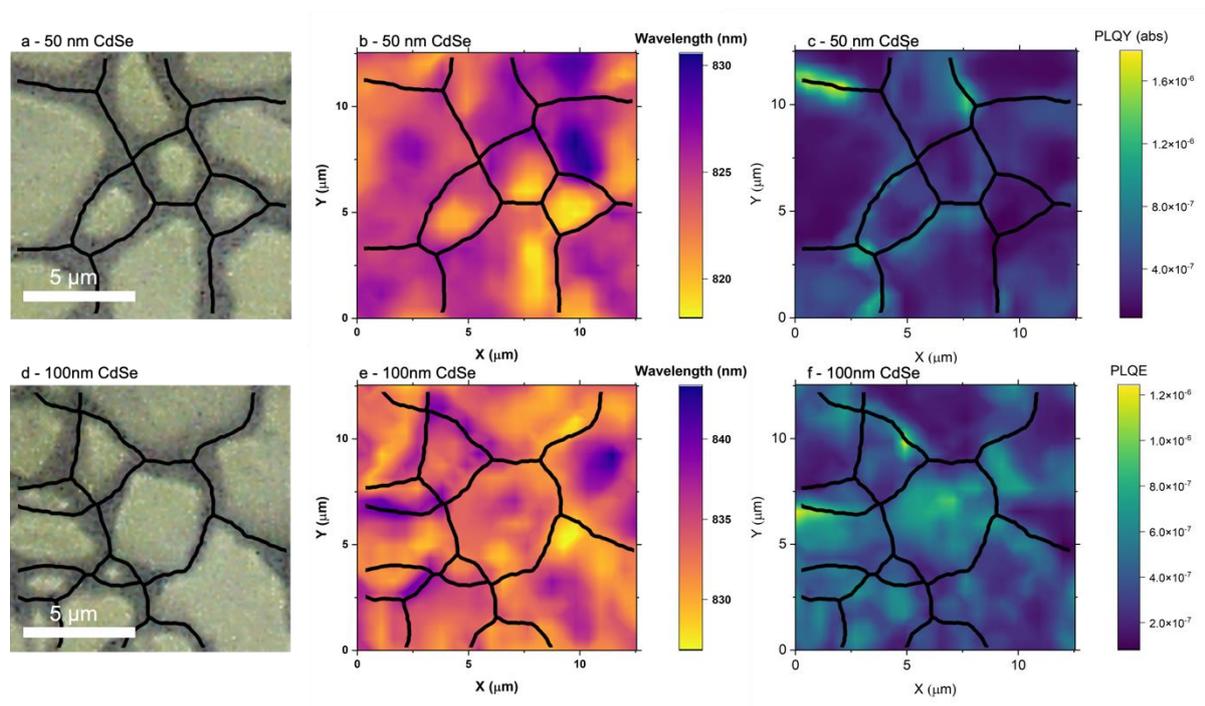

*Figure S11. a), b) and c)/d), e) and f) present white light reflection, peak wavelength and peak PLQY plots (±20 nm about the central peak) for Cl-treated 50 nm CdSe and 100 nm Cl-treated CdSe samples. We note that the same colour scales are used for different ranges in b)/e) and c)/f) to allow for better visualisation of results. As a guide to the eye, black lines present approximate grain boundaries.*

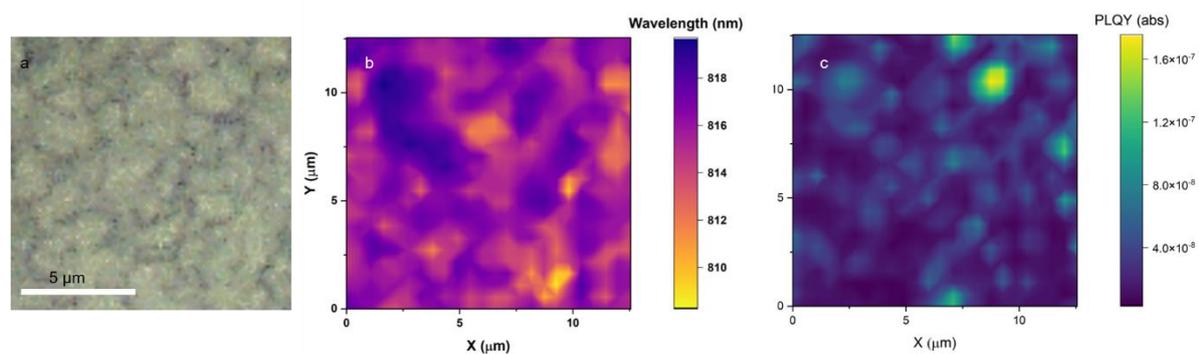

*Figure S12. a), b) and c) present white light reflection, peak wavelength and peak PLQY plots (±20 nm about the central peak) for untreated 0 nm CdSe.*

*Supporting Information Note 9 – overlaid maps*

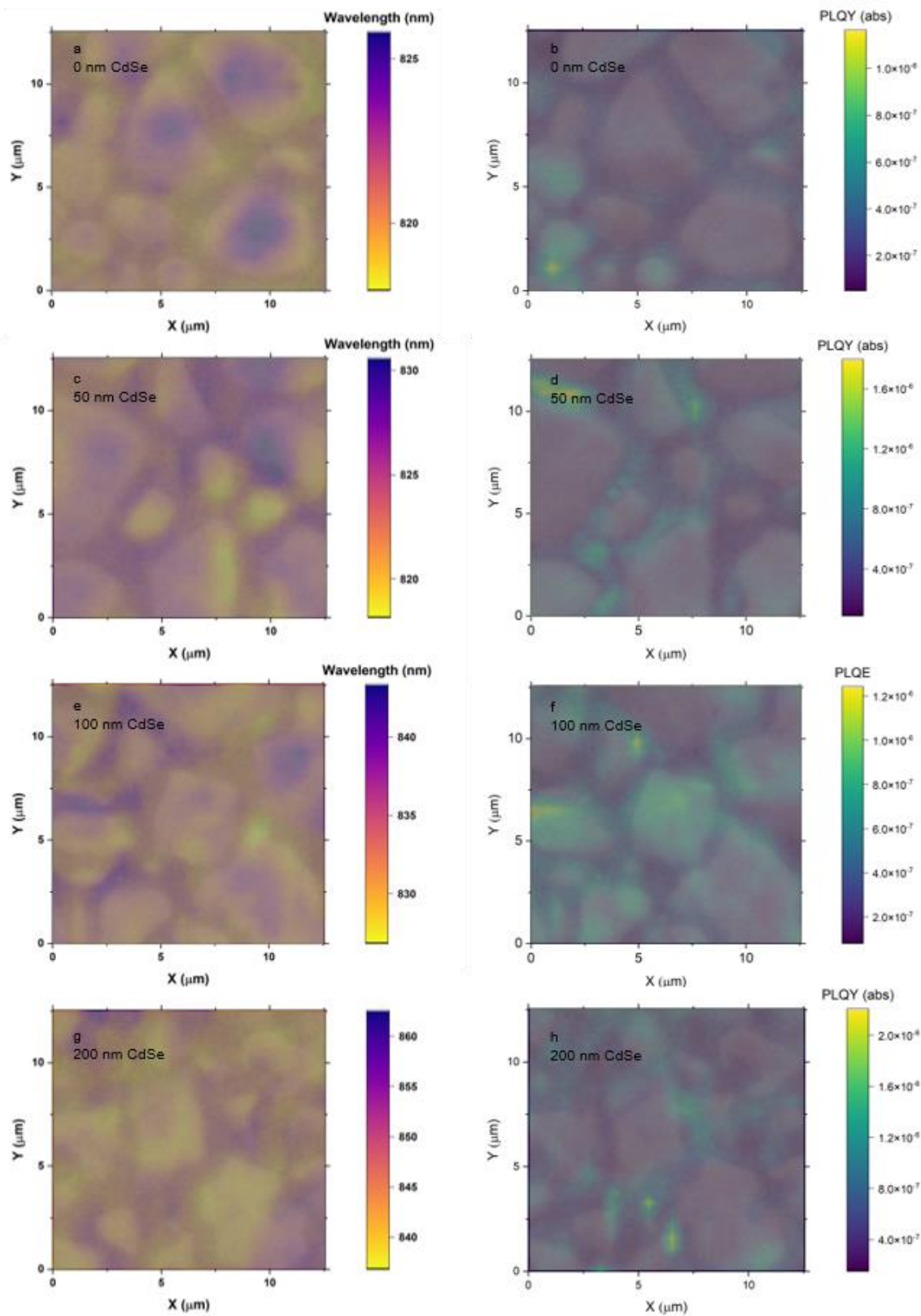

*Figure S13. a)/c)/e)/g) and b)/d)/f)/h) present peak wavelength maps and PLQE maps overlaid with the white light sample reflection images for Cl-treated 0 nm/50 nm/100 nm/200 nm CdSe.*

*Supporting Information Note 10 – correlation maps*

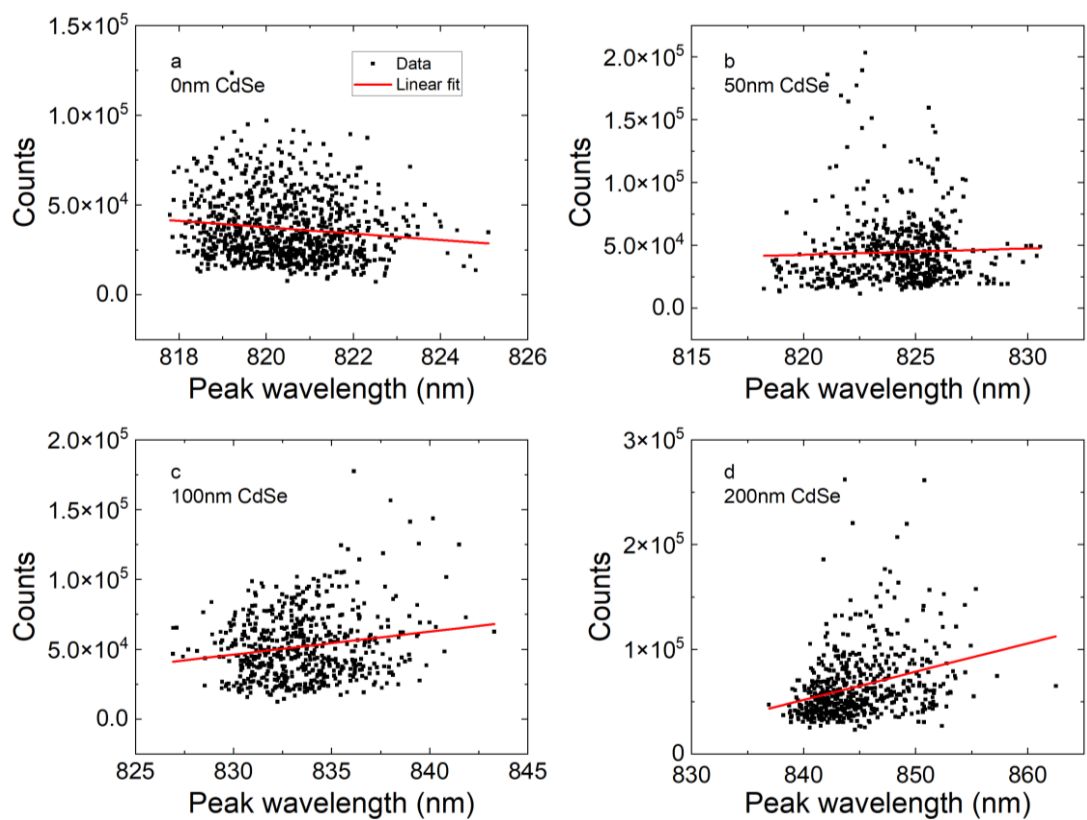

*Figure S14. a)/b)/c)/d) present correlation maps between peak wavelength and total PL counts for Cl-treated 0 nm/50 nm/100 nm/200 nm CdSe.*

*Supporting Information Note 11 – plots of long-wavelength PLQY*

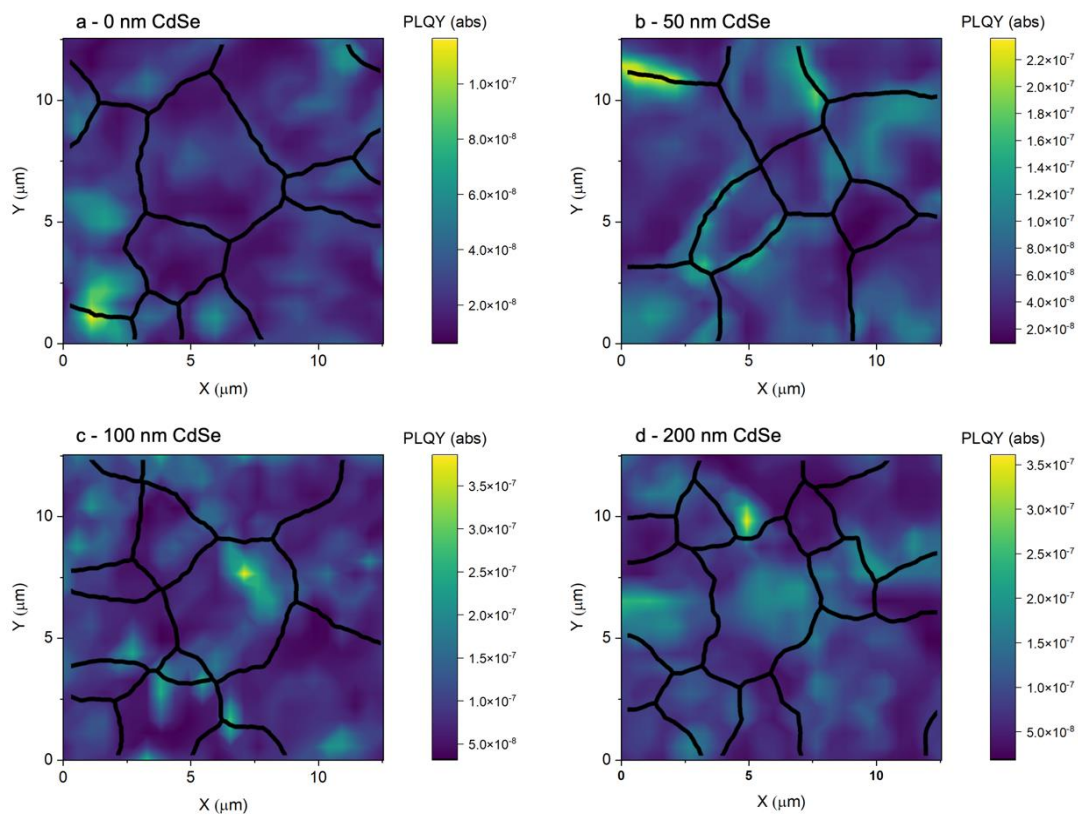

*Figure S15. a)/b)/c)/d) presents long wavelength PLQYs for Cl-treated 0 nm/50 nm/100 nm/200 nm CdSe. PLQYs are for wavelength regions of 870 nm-950 nm/870 nm-950 nm/890 nm-950 nm/900 nm-950 nm respectively (chosen to avoid the edge of the main PL peak). As a guide to the eye, black lines present approximate grain boundaries.*